\documentclass[a4paper,fleqn,usenatbib]{mnras}

\usepackage{newtxtext,newtxmath}

\usepackage[T1]{fontenc}
\usepackage{ae,aecompl}

\usepackage{graphicx}	
\usepackage{amsmath}	
\usepackage{amssymb}	

\usepackage[usenames,dvipsnames]{color}
\definecolor{MyDarkBlue}{rgb}{0,0.1,0.7}
\hypersetup{pdfborder={0 0 0},colorlinks,breaklinks=true,
  urlcolor={MyDarkBlue},citecolor={MyDarkBlue},linkcolor={MyDarkBlue} }
\usepackage{hyperref,graphicx}
\usepackage{rotating}
\usepackage[export]{adjustbox}
\usepackage{setspace}
\usepackage{booktabs}

\usepackage{layouts}

\DeclareMathOperator*{\argmax}{arg\,max}

\newcommand{\first}{1\textsuperscript{st}}
\newcommand{\second}{2\textsuperscript{nd}}
\newcommand{\third}{3\textsuperscript{rd}}
\newcommand{\fourth}{4\textsuperscript{th}}
\newcommand{\fifth}{5\textsuperscript{th}}
\newcommand{\sixth}{6\textsuperscript{th}}

\newcommand{\kth}[1]{\textit{#1}\,\textsuperscript{th}}

\newcommand{\mybold}{\boldsymbol}
\newcommand{\pir}{{\mathrm \pi}}

\newcommand{\pgiven}{\,|\,}

\newcommand{\per}{P}
\newcommand{\perdot}{\dot{\per}}
\newcommand{\pspace}{\per-\perdot}
\newcommand{\afreq}{\Omega}
\newcommand{\afreqdot}{\dot{\afreq}}
\newcommand{\cage}{\tau_{\rm c}}
\newcommand{\bfield}{B}
\newcommand{\bfieldQED}{\bfield_{\rm QED}}
\newcommand{\bfieldLC}{\bfield_{\rm LC}}
\newcommand{\sdpow}{L_{\rm sd}}
\newcommand{\xlum}{L_{\rm X}}

\newcommand{\pradLC}{R_{\rm LC}}

\newcommand{\tvel}{v_{\rm T}}
\newcommand{\tvelM}{\overline{\tvel}}
\newcommand{\tvelWM}{\overline{\tvel^{*}}}
\newcommand{\tvelS}{\sigma_{\tvel}}
\newcommand{\tvelWS}{\sigma^{*}_{\tvel}}

\newcommand{\ktbb}{T_{\rm bb}}
\newcommand{\ktbbM}{\overline{\ktbb}}
\newcommand{\ktbbWM}{\overline{\ktbb^{*}}}
\newcommand{\ktbbS}{\sigma_{\ktbb}}
\newcommand{\ktbbWS}{\sigma^{*}_{\ktbb}}

\newcommand{\dset}{\mybold{X}} 
\newcommand{\dpoint}[1][]{\ifthenelse{\equal{#1}{}}{\mybold{x}}{\mybold{x_{\rm #1}}}} 
\newcommand{\pset}{\mybold{\theta}}

\newcommand{\lset}{\mybold{Z}} 
\newcommand{\lpoint}[1][]{\ifthenelse{\equal{#1}{}}{z}{z_{\rm #1}}}
\newcommand{\maxclust}{m_{\rm max}}

\newcommand{\gmean}[1][]{\ifthenelse{\equal{#1}{}}{\mybold{\mu}}{\mybold{\mu_{\rm #1}}}}
\newcommand{\gcovar}[1][]{\ifthenelse{\equal{#1}{}}{\mybold{\Sigma}}{\mybold{\Sigma_{\rm #1}}}}
\newcommand{\gweight}[1][]{\ifthenelse{\equal{#1}{}}{\mybold{w}}{w_{\rm #1}}} 

\newcommand{\dparams}[1][]{\ifthenelse{\equal{#1}{}}{\mybold{\alpha_{\rm #1}}}{\alpha_{\rm #1}}}
\newcommand{\dparam}{\alpha}
\newcommand{\gauss}{\mathcal{N}}

\newcommand{\niwishart}{\mathcal{NIW}}
\newcommand{\niwmean}{\mathcal{\mybold{\mu_{\rm 0}}}}
\newcommand{\niwsfac}{\beta}
\newcommand{\niwsmat}{\mybold{\psi}}  
\newcommand{\niwdof}{\nu} 
\newcommand{\dbase}{\mathcal{H}}

\title[Classification of pulsars with DPGMM]
{Classification of pulsars with Dirichlet process Gaussian mixture model}

\author[F.\ Ay, G.\ {\.I}nce, M.~E.\ Kama\c{s}ak, K.~Y.\ Ek\c{s}i]{
Fahrettin Ay,$^{1}$\thanks{E-mail: \href{mailto:ayf@itu.edu.tr}{ayf@itu.edu.tr}}
G{\"o}khan {\.I}nce,$^{1}$
Mustafa E.\ Kama\c{s}ak$^{1}$ 
and K.~Yavuz Ek\c{s}i$^{2}$\thanks{E-mail: \href{mailto:eksi@itu.edu.tr}{eksi@itu.edu.tr}}
\\
$^{1}$Istanbul Technical University, Faculty of Computer and Informatics, Computer Engineering Department, 34469,  \.Istanbul, Turkey\\
$^{2}$Istanbul Technical University, Faculty  of Science  and  Letters,  Physics Engineering  Department, 34469,  \.Istanbul, Turkey
}

\pubyear{2019}

\begin{document}
\label{firstpage}
\pagerange{\pageref{firstpage}--\pageref{lastpage}}
\maketitle

\begin{abstract}
Young isolated neutron stars (INS) most commonly manifest themselves as rotationally powered pulsars (RPPs) which involve conventional radio pulsars as well as gamma-ray pulsars (GRPs) and rotating radio transients (RRATs). 
Some other young INS families manifest themselves as anomalous X-ray pulsars (AXPs) and soft gamma-ray repeaters (SGRs) which are commonly accepted as magnetars, i.e.\ magnetically powered neutron stars with decaying super-strong fields. 
Yet some other young INS are identified as central compact objects (CCOs) and X-ray dim isolated neutron stars (XDINSs) which are cooling objects powered by their thermal energy. 
Older pulsars, as a result of a previous long episode of accretion from a companion, manifest themselves as millisecond pulsars and more commonly appear in binary systems. 
We use Dirichlet process Gaussian mixture model (DPGMM), an unsupervised machine learning algorithm, for analyzing the distribution of these pulsar families in the parameter space of period and period derivative. 
We compare the average values of the characteristic age, magnetic dipole field strength, surface temperature and  transverse velocity of all discovered clusters. 
We verify that DPGMM is robust and provides hints for inferring relations between different classes of pulsars. 
We discuss the implications of our findings for the magneto-thermal spin evolution models and fallback discs.  
\end{abstract}




\section{Introduction}
\label{intro}


Rotationally powered pulsars (RPPs), the most common manifestation of neutron stars, were discovered first as radio pulsars \citep{hew68}. 
More than $\sim 2600$ RPPs discovered to date are understood to be strongly magnetized ($\bfield \sim 10^{12}$~G), rapidly rotating (with spin periods $P \sim 0.1$~s) neutron stars spinning down by torques due to magnetic dipole radiation and particle emission. 
These objects emit most of the energy, tapped from their rotational kinetic energy, in X- and gamma-ray bands. 
The total radiative luminosity of RPPs is thus below their spin-down power $\sdpow \equiv -I \afreq \afreqdot$ where $I$ is the moment of inertia of the compact object, $\Omega=2\pir/P$ is the spin angular frequency and $\afreqdot$ is its time derivative \citep[see][for a review]{man17}. 
The bulk of the RPP population are the classical radio pulsars with periods $\per = 0.2-2$~s, period derivatives $\perdot = 10^{-16} - 10^{-13}$~s~s$^{-1}$, magnetic fields $\bfield \sim 10^{10}-10^{13}$~G and characteristic ages $\cage \equiv \per/2\perdot \sim 10^3 - 10^7$~years. 
Some of the RPPs are associated with supernova remnants (SNRs). 
Young RPPs are usually isolated objects in the sense that they have no binary companion. 
The pulsations are attributed to the beaming of the emitted radiation and the ``lighthouse'' effect produced by the spin of the object. 
Thus, apart from distant and dim objects, there could be many RPPs whose beam does not sweep our line of sight.

A population among the RPPs are the millisecond pulsars \citep[MSP;][]{bac+82} which are old but ``recycled'' pulsars \citep{alp+82}; 
these objects are understood to descent from low-mass X-ray binaries \citep[LMXBs; see][for reviews]{bha91,tau06} where the neutron star 
spins-up by accreting matter from a companion via a disc. 
The accreted matter transfers angular momentum to the neutron star thereby 
spinning it up to millisecond periods and the magnetic field of the neutron star is reduced to values $B < 10^{9}$~G in the process.

More than 200 $\gamma$-ray pulsars (GRPs) discovered
 by the Large Area Telescope onboard Fermi \citep[Fermi-LAT;][]{atw+09} constitute a subclass of RPPs \citep{abd+10,abd+13}. 
 All known GRPs have $\sdpow > 10^{33}$ erg s$^{-1}$. 
 About half of the GRPs detected by Fermi are radio-quite. 
 This possibly indicates that the ``radio beam'' which is narrower than the ``gamma beam'' does not pass from our line of sight. 
 Approximately half of the discovered GRPs are MSP which was unexpected before the discovery \citep{abd+09msec}.

Another recently discovered family of RPPs is the rotating radio transients (RRATs) identified in the Parkes multibeam survey \citep{mcl+06}. 
Unlike ordinary pulsars detected by searches in the frequency domain, RRATs show sporadic emission detected through their bright single pulses. 
These objects show bursts of duration $2-30$~ms with an interval in the range 4 min-3 hr. 
Their spin periods are in the range $0.4-7$~s. 
The period derivatives are measured in 3 sources implying magnetic fields $B \sim 10^{12}-10^{14}$~G and characteristic  ages $\cage \sim 0.1-3$~Myr. 

In the last two decades many young isolated neutron star (INS) families \citep[see][for reviews]{pop08,kas10,har13,saf17} other than radio pulsars are identified. 
These objects are usually radio-quiet and have X-ray luminosities exceeding the spin-down power of the compact object. 
This indicates the availability of energy budgets other than rotational kinetic energy and the possibility of evolutionary paths other than those leading to RPPs. 

Of these groups of objects soft gamma ray repeaters (SGRs) and anomalous X-ray pulsars (AXPs) are commonly assumed to be magnetars (see \citet{woo06,mer08,rea14,tur+15,mer+15,kas17,gou18} for reviews) i.e.\ INSs with large dipole magnetic fields $\bfield \sim 10^{14}-10^{15}$~G as inferred from their rapid spin down $\perdot\sim 10^{-13}-10^{-11}$~s~s$^{-1}$ \citep{kou+98} and slow periods clustered between $\per = 2-12$ seconds.  
According to the magnetar model \citep{dun92,tho96} the persistent X-ray emission of these objects with X-ray luminosity of $\xlum\sim 10^{35}-10^{36}$~erg~s$^{-1}$ is powered by the decay of this strong magnetic field \citep{pac92,tho95} in excess of the quantum critical limit $\bfieldQED = 4.4\times 10^{13}$~G. 
These objects occasionally show super-Eddington outbursts and, very rarely, giant bursts which are addressed in the magnetar model with the breaking of the neutron star crust due to magnetic stresses and reconfiguration of their fields, respectively. 
The magnetars are young objects as implied by their characteristic ages $\cage \sim 10^4$~years and about half of them being associated with SNRs. 

X-ray dim isolated neutron stars (XDINSs), or sometimes called ``magnificent seven'', are the 7 nearby neutron stars identified through their thermal X-ray emission \citep{oze13,pot+15,mer11} with luminosities of order $L_{\rm X} \sim 10^{30}-10^{32}$~erg~s$^{-1}$ \citep[see][for reviews]{hab07,kap08,tur09}. 
They have a period range similar to the AXP/SGR family \citep{ham+17}, but are typically older, with characteristic ages $\cage \sim 10^{5}-10^6$~years and kinematic ages of a few $10^6$~years \citep{tet+10,tet+11,tet+12}. 
They have inferred dipole magnetic fields of $B \sim 10^{13}$~G, an order of magnitude lower than magnetars and an order of magnitude higher than conventional pulsars, but their surface magnetic fields as inferred from the narrow absorption features are 7 \citep{bor+15}  and 5 \citep{bor+17} times larger in the case of RX J0720.4--3125 and RX J1308.6+2127, respectively.

Yet another young NS family is the central compact objects  \citep[CCOs; see][for a review]{deL17}  in supernova remnants (SNRs). 
These $\sim 10$ objects show no sign of RPP activity. 
Their X-ray spectra is dominated by the thermal emission showing some similarities with magnetars yet they are two orders of magnitude less luminous ($\sim 10^{33}- 10^{34}$~erg s$^{-1}$). 
The periods measured from 3 CCOs are in the range $0.1-0.4$~s and the measured period derivatives imply that the dipole fields of these objects are in the range $B\sim 10^{10}-10^{11}$~G \citep{got07,got09,hal11}, and hence sometimes are called ``anti-magnetars'' \citep{hal10,got+13}. 
Yet there is evidence that these objects have much stronger ``hidden'' magnetic fields \citep{vig12,vig+13,tor+16} as implied by the highly anisotropic emission \citep{sha11} leading to the observed high pulsed fraction. 
According to the ``field burial scenario'' \citep{mus95,you95,gep+99,ho11,ber+13,igo+16} this is due to an initial fallback accretion episode the nascent neutron star had suffered soon after the SN explosion. 
An exception among the CCO family is 1E~161348--5055 in SNR RCW 103 which has an unusually long period of 6.7 hours \citep{deL08} and has recently shown magnetar-like bursts \citep{rea+16}. 

Recent discoveries are blurring the borders of the classes: (i) The existence of high magnetic field RPPs (\citet{piv00}; see \citet{ng11} for a review) that have magnetic field strengths comparable and, in some cases, exceeding that of some magnetars and some of which has even shown magnetar-like X-ray bursts \citep{gav08,arc+16,gog+16}; (ii) radio detection from some magnetars \citep{cam06,cam07a,lev10}; (iii) identification of low-magnetic field magnetars \citep{rea10} i.e.\ INS showing magnetars bursts but with ordinary inferred dipole field strengths \citep[see][for a review]{tur13}.
These discoveries  indicate that the \textit{dipole} field strength is not the single parameter leading to the different manifestations favouring the early suggestions \citep{gav02,eks03,ert03,mcl03} that what causes the magnetar activity could be in the higher multipoles \citep{alp11,rod+16}.

In fact there are other observations suggesting that AXP/SGRs have ``low'' \textit{dipole} fields:
(i) They are not detected in the Fermi/LAT observations \citep{sas10,abd10} in the GeV range though, according to the outer gap model \citep{zha97}, they are expected \citep{che01} to emit high energy gamma-rays 
should they have super-strong magnetic dipole fields $\gtrsim 10^{14}$~G \citep{ton11}. 
(ii) their transverse velocities are measured \citep{hel07,del12,ten+12,ten+13} to be $200\pm 100$~km~s$^{-1}$, similar to the velocities of RPPs though they are expected to have exceptionally large space velocities, $\sim 1000$~km~s$^{-1}$, due to getting stronger kicks \citep{dun92,tho93} via the rocket propulsion effect should they have super-strong dipole fields. 
(iii) They do not commonly show the signature of the expected strong energy
injection in the SNRs due to hosting a neutron star with a birth period of milliseconds
 \citep{vin06,mar+14,bor17} (but see \citet{tor17}) that could produce super-strong fields by dynamo action \citep{dun92,tho93}.
		
Given the transitivity among the INS families the question naturally arises whether the classification of INS above is robust. 
Could we expect, for example, that more of the RPPs to show magnetar-like behaviour, or that more magnetars to appear in the radio band? What parameters could be leading to the different manifestations for objects with similar period and period derivatives? The magneto-thermal evolutionary theory \citep[e.g.][]{vig12,vig+13} addresses this diversity by the presence of toroidal magnetic fields hidden in the crust as an extra parameter shaping the lives of pulsars. 
The fallback disc model \citep{alp01,cha+00} invokes the mass and specific angular momentum \citep{ert+07,ert+09} of a putative supernova fallback disc as two parameters leading to the diversity.



A well-known unsupervised machine learning algorithm, Gaussian mixture model 
\citep[GMM;][]{press+07},
 had been used  by \citet{lee+12} to analyze the distribution of pulsars on the $\pspace$ diagram and identified six Gaussian clusters (2 for millisecond---recycled---pulsars and 4 for young pulsars). 
\citet{igo13} showed that GMM is over-sensitive to the data and does not demonstrate robust clustering performance. 
There are more advanced variations of GMM such as Dirichlet process Gaussian mixture model (DPGMM). 
This model has the advantage over GMM that the number of distributions are determined automatically though it is not used as widely as GMM especially by the astronomy community. 
In addition its generalization performance is better than GMM, 
as Bayesian methods are more robust against overfitting \citep{ras00,hel08, wit+16}. 
We have recently showed that DPGMM is better than GMM in both clustering and classifying pulsars in the $\pspace$ parameter space \citep{ay+19}.

The organization of our paper is as follows. 
In \S~\ref{sec:method} we review DPGMM employed in this work.
In \S~\ref{sec:results} we present our results and finally, in \S~\ref{sec:discuss}, we discuss the implications of our findings.

\section{Method}
\label{sec:method}

\subsection{Dirichlet process Gaussian mixture model}
\label{subsec:dpgmm_method}

Here we will briefly sketch the method employed in this work. The detailed information for DPGMM can be found in \citep{shi2+09, del+18}.  

In finite mixture models, it is assumed that the data consists of a certain number of clusters in which each cluster is generated by a probability distribution. 
Let $\dset^{\rm T} = \lbrace\dpoint[j]\rbrace_{\rm j=1}^{\rm N}$ 
be the collection of $d$-dimensional observed data with $N$ number of instances and assume further that there are $m$ distributions that generate $\dset$. 
Because it is not known by which  distribution a certain data point is generated, it is useful to define a latent data $\lset$ for distribution assignments; $\lset^{\rm T} = \lbrace\lpoint[j]\rbrace_{\rm j=1}^{\rm N}$ 
where $\lpoint[j]=i$ means that $\dpoint[j]$ is generated by the \kth{i} distribution,   
$j \in \lbrace 1, 2, \ldots, N \rbrace$ 
and 
$i \in \lbrace 1, 2, \ldots, m \rbrace$.  

The joint probability of a data point and \kth{i} distribution is defined as 
\begin{equation}
\label{eq:joint_probability}
p(\dpoint[j], \lpoint[j]=i) = p(\dpoint[j] \pgiven\lpoint[j]=i) \, p(\lpoint[j]=i) \enskip .
\end{equation}
The joint probability value indicates how likely a data point belongs to a cluster. 
Because $\dpoint[j]$ must be populated by one of the $m$ distributions, we can obtain marginal probability of $\dpoint[j]$ from equation \eqref{eq:joint_probability} as follows 
\begin{equation}
\label{eq:marginal_probability}
p(\dpoint[j]) = \sum_{i=1}^{m}{p(\dpoint[j] \pgiven\lpoint[j]=i) \, p(\lpoint[j]=i)} \enskip .
\end{equation}
The marginal probability value can be considered as the probability that a data point belongs to \textit{any} cluster. 
On the right side of the equation, the second term is the probability of the \kth{i} distribution which is also known as the weight of the \kth{i} cluster such that $\gweight[i] \equiv p(\lpoint[j]=i)$ where $\sum_{i=1}^{m}{\gweight[i]}= 1$. 
Besides, if we assume that each cluster is distributed normally, then the first term on the right side of the equation is defined as Gaussian distribution;
\begin{equation}
\label{eq:normal_dist}
\gauss\left(\dpoint[j] \pgiven \gmean[i], \gcovar[i] \right) = 
	\frac{\exp \left(-\frac{1}{2}(\dpoint[j]-\gmean[i])^{\rm T}\,\gcovar[i]^{-1}(\dpoint[j]-\gmean[i]) \right)}
	{(2\pir)^{d/2} \lvert \gcovar[i] \rvert^{1/2}} \enskip ,
\end{equation}
where $\gmean[i]$ is $d$-length mean vector and $\gcovar[i]$ is $d \times d$ covariance matrix of the \kth{i} Gaussian distribution. 
In this case, the finite mixture model is called as Gaussian mixture model (GMM).   

In Bayesian approach, GMM parameters  
$\pset = \lbrace \gweight, \gmean, \gcovar \rbrace$
are also modeled with probability distributions.
The choice of distribution families is based on \textit{conjugate prior} principle \citep{rai61} for the computational simplicity. 
Accordingly, the prior distributions for the parameters of Gaussian distributions are selected as Normal-Inverse-Wishart distribution;
\begin{equation}
\label{eq:niw_prior}
\lbrace \gmean, \gcovar \rbrace \sim 
\niwishart \left(\niwmean, \niwsfac, \niwsmat, \niwdof\right) \enskip , 
\end{equation}
where $\lbrace \niwmean, \niwsfac, \niwsmat, \niwdof  \rbrace$ are mean, scaling factor, scale matrix and degrees of freedom. Furthermore, if we assume that the cluster weights are Dirichlet process \citep[DP;][]{ferg73} distributed  
\begin{equation}
\label{eq:dp_prior}
\gweight \sim DP(\dparam, \dbase) \enskip ,
\end{equation}
then the mixture model is called as Dirichlet process Gaussian mixture model (DPGMM). In the equation, $\dparam$ is the concentration parameter and $\dbase$ is the base distribution for DP. 
Because a realization from DP is a infinite length probability vector, DPGMM is a non-parametric, i.e.\ infinite, mixture model. 

In this study, we employed \textit{scikit-learn} \citep{sklearn} implementation of DPGMM that uses variational inference \citep{att00, bis06, ble06} for learning model parameters. We also benefited from many open source software for scientific computations, analyzes and visualizations, such as 
\textit{scipy} \citep{scipy2}, \textit{numpy} \citep{numpy}, \textit{pandas} \citep{pandas}, \textit{matplotlib} \citep{matplotlib}, \textit{seaborn} \citep{seaborn_org} and \textit{scikit-image} \citep{skimage}. 
In addition, 
\textit{Spyder}\footnote{\url{https://www.spyder-ide.org/}} 
and 
\textit{IPython} \citep{ipython} were used as development environments. 

\subsection{Data Collection}
\label{subsec:data_collection}

We examine pulsar distribution on $\pspace$ parameter space in logarithmic scale. In order to increase data samples as much as possible, various neutron star catalogs are combined.  
Most of the pulsar data are obtained from 
Australia Telescope National Facility (ATNF) Pulsar Catalog\footnote{\url{http://www.atnf.csiro.au/research/pulsar/psrcat/}} that contains up to 2658 instances with numerous source types. 
In addition, 30 instances of SGRs and AXPs (either confirmed or candidates) from McGill Online Magnetar Catalog\footnote{\url{http://www.physics.mcgill.ca/\~pulsar/magnetar/main.html}} \citep{ola14}
and 40 instances of other thermally emitting neutron stars such as CCOs and XDINSs from \citep{vig+13}\footnote{\url{http://www.neutronstarcooling.info/}} are collected. 
Additionally, 107 RRATs from RRATOLOG\footnote{\url{http://astro.phys.wvu.edu/rratalog/}} \citep{mcl+06,kea+10}. 
and 117 GRPs from \textit{The Second Fermi-LAT} catalogue\footnote{\url{https://heasarc.gsfc.nasa.gov/W3Browse/fermi/fermil2psr.html}} \citep{abd+13} are included. 
Finally, a recently discovered \citep{tan+18} RPP with an exceptional long period is also included to the data set. 
As a result, we obtained an amount of 2166 INSs with the observed spin parameters by combining all of these data sources.      

\section{Results}
\label{sec:results}
 
\subsection{Application of DPGMM}
\label{subsec:dpgmm_application}

DPGMM has a relatively large number of hyper-parameters to be fixed. 
For example, although DPGMM is an infinite mixture model, \textit{scikit-learn} implementation of DPGMM requires the hyper-parameter $\maxclust$ that corresponds to the \textit{maximum} number of clusters to be set to an appropriate value as it would not be practical to consider infinite number of clusters during fitting. 
We have set the value of $\maxclust$ as 10 as we did not expect to find more than 10 clusters given the number of pulsar families and the 6 clusters already determined by GMM \citep{lee+12}. 

Another hyper-parameter is the concentration parameter $\dparam$ that determines how sparse distributions are allowed to be and thus may effect the number of clusters the model can find. 
For a given $\dparam$, the optimum number of clusters that best represent the data among the $\maxclust$ clusters and their parameters are determined by DPGMM during the fitting process. 
We have seen that the value of $\dparam$ in the literature varies significantly: e.g.\  
\citet{hai13} set $\dparam=0.01$ while \citet{shi2+09} set $\dparam=1$. 
In another example, \citet{che+15} have searched optimum value for $\dparam$ in a range between $0.001$ and $100$, and observed that the results are relatively same independent of different $\dparam$ values.	

We created DPGMMs for different $\dparam$ values within a wide range from $10^{-10}$ to $10^{10}$.  
As a result, we have observed that DPGMM for this wide range of $\dparam$ values discovered either 6 or 7 clusters depending on whether $\dparam$ is lower or higher than $10^4$, respectively.
We have also observed that the estimated model parameters are remarkably independent of $\dparam$ for all models that discovered 6 clusters 
\textcolor{blue}{($\dparam \leq 10^4$).} The same situation is also true for all models that discovered 7 clusters ($\dparam>10^4$). 
In short, we have obtained two different stable models; one having 6 clusters and the other 7 clusters. 
The recommended value of $\dparam$ is much less than 1 and for such values our model finds 6 clusters.   
We have employed Bayesian information criteria \citep{schwarz78} and Akaike information criteria \citep{aka74} in a former study \citep{ay+19} to determine the optimum values of the clusters in which we found 6 clusters. This result also is consistent with the number of clusters found by \citet{lee+12} who used multidimensional Kolmogorov-Smirnov test.
We thus selected the model with 6 clusters as the optimal one and our results with 7 clusters are not shown here. 
In \autoref{table:dpgmm_components_params}, the estimated parameters of DPGMM created with $\maxclust=10$ and $\dparam=10^{-5}$ are shown.

%
\begin{table}
\centering
\begin{tabular}{@{}|c|c|c|c|@{}}
\toprule
\multicolumn{4}{|c|}{\textbf{DPGMM Parameters}} \\ \midrule

\textbf{Cluster} & $\gweight[~]$ & $\gmean$ & $\gcovar$  \\ \midrule
\first  & 0.0787 & $\begin{bmatrix} -2.3782 & -19.8306 \end{bmatrix}$ & $\begin{bmatrix} 0.0731 & 0.1019 \\ 0.1019 & 0.4127 \end{bmatrix}$ \\ \midrule
\second  & 0.0384 & $\begin{bmatrix} -1.5694 & -18.4996 \end{bmatrix}$ & $\begin{bmatrix} 0.2776 & 0.2627 \\ 0.2627 & 0.9198 \end{bmatrix}$ \\ \midrule
\third  & 0.3344 & $\begin{bmatrix} -0.0148 & -15.3005 \end{bmatrix}$ & $\begin{bmatrix} 0.1303 & 0.2647 \\ 0.2647 & 0.9584 \end{bmatrix}$ \\ \midrule
\fourth  & 0.4111 & $\begin{bmatrix} -0.2817 & -14.5136 \end{bmatrix}$ & $\begin{bmatrix} 0.0662 & 0.0536 \\ 0.0536 & 0.4303 \end{bmatrix}$ \\ \midrule
\fifth  & 0.1129 & $\begin{bmatrix} -0.7303 & -13.5903 \end{bmatrix}$ & $\begin{bmatrix} 0.1145 & 0.0096 \\ 0.0096 & 0.5683 \end{bmatrix}$ \\ \midrule
\sixth  & 0.0244 & $\begin{bmatrix} 0.3350 & -12.2949 \end{bmatrix}$ & $\begin{bmatrix} 0.1864 & 0.4824 \\ 0.4824 & 2.1642 \end{bmatrix}$ \\ \midrule

							
\end{tabular}
\caption{The estimated parameters of DPGMM ($\maxclust=10$, $\dparam=10^{-5}$). ~$\gweight[~]$, $\gmean$ and $\gcovar$ indicate weight, location and shape of the corresponding Gaussian distribution, respectively.}
\label{table:dpgmm_components_params}
\end{table}
%

\subsection{Parameter space classification}
\label{subsec:pspace_classification}

\begin{figure*}
  \includegraphics[width=\linewidth,height=\linewidth,center]{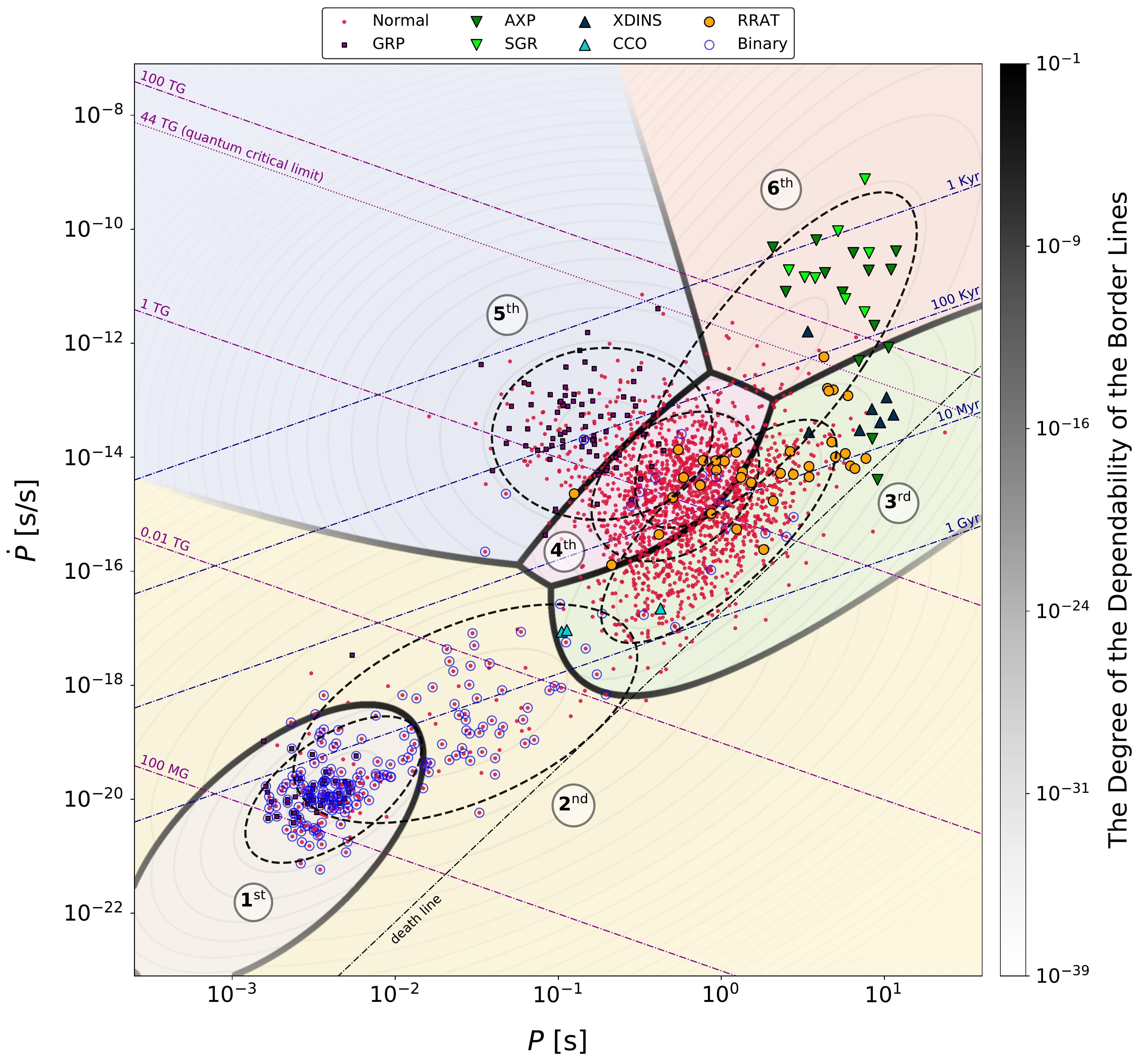}
\vskip -2mm
  \caption{The $\pspace$ parameter space classified by DPGMM. Each cluster region is uniquely enumerated and colored. The decision boundaries are illustrated with thick solid lines of varying shades representing the degree of the dependability of the border lines (see \S~\ref{subsec:pspace_classification}). Bivariate Gaussian curves are represented with $2\sigma$ confidence ellipses. }
  \label{fig:dpgmm_pp1}
\end{figure*}

In \autoref{fig:dpgmm_pp1}, all clusters and decision boundaries on $\pspace$~ parameter space revealed by DPGMM are illustrated. Bivariate
Gaussian curves are represented by 2$\sigma$ confidence ellipses and each cluster region is enumerated and colored uniquely in here.
The mixture models are \textit{inductive} models so that after fitting a model to a data set, it is able to predict the cluster in which a `novel' instance should belong. 
Formally, the cluster label of a novel instance $\dpoint[j+1]$ is determined as  
\begin{equation}
\label{eq:cluster_assignment}
\lpoint[j+1] = \argmax_{i}{p(\dpoint[j+1], \lpoint[j+1]=i)} \enskip .
\end{equation}  
%
If we classify the entire parameter space with this approach, the separating line between cluster regions naturally arise as shown in \autoref{fig:dpgmm_pp1} and in \autoref{fig:BLC-Edot}. These lines are called as decision boundaries such that the cluster assignment by the model is different for the one side of the line and the other side. 

In the case of mixture of Gaussian distributions, there is a possibility of being more than one region for a single cluster on classified parameter space. 
To illustrate, there are two regions for the \sixth\ cluster in \autoref{fig:BLC-Edot}; one is on the top left, and the other is on the bottom right. 
It is significant that all pulsars belonging to the \sixth\ cluster are in the top left region while the bottom one does not cover any pulsars at all. 
Thus, it can be said that instead of the \sixth\ cluster, the bottom region should be merged with the neighbor region corresponding to the \second\ cluster. 
Therefore, the separating line between the \second\ cluster and the bottom region of the \sixth\ cluster seems incorrect. 
For this reason, we examined the \textit{dependability} of boundary lines in this study. 

As mentioned before, the joint probability defined in equation \eqref{eq:joint_probability} and used in equation \eqref{eq:cluster_assignment} for cluster assignments shows that how likely a data point $\dpoint[j]$ is a member of \kth{i} cluster. 
Its value depends on the cluster weight and the multivariate Gaussian distribution defined in equation \eqref{eq:normal_dist} where the exponent term in the numerator is actually half the negative of the square of Mahalanobis distance \citep{mah36,de+00} that is a distance measure in terms of standard deviation. 
Roughly speaking, it can be said that the cluster assignment for a data point is determined by the weight of clusters and the Mahalanobis distance between the data point and cluster means. 
The cluster assignment is predominately determined by the weight of the clusters for nearby data points.
On the other hand, the effect of Mahalanobis distance on the decision making of the model increases for the receding data points.
As the Mahalanobis distance is not a deterministic measure, the cluster assignments may be incoherent for the distant regions. 
To illustrate, the bottom right region in \autoref{fig:BLC-Edot} is separated from the \second\ cluster and assigned to the \sixth\ cluster by the model even though all pulsars belonging to the \sixth\ cluster are in top left region. 
Therefore, we defined the dependability of the cluster assignments as the marginal probability values specified in equation \eqref{eq:marginal_probability} that can be considered as how likely a data point is a member of any cluster, as stated before. 
If the marginal probability of $\dpoint[j]$ is lower, then the dependability of the cluster assignment for this data point is not reliable as well. 
In this way, we showed the \textit{degree of the dependability} of border lines in \autoref{fig:dpgmm_pp1} and \autoref{fig:BLC-Edot} based on the reliability of data points over border lines, so that the border line is darker where it is more reliable, and it is fainter where it is less reliable.
Accordingly, it is obvious that the color of the border line between the \second\ and the \sixth\ clusters in \autoref{fig:BLC-Edot} is very faint that means it is not reliable as expected. 
In this way,  we tried to eliminate the handicap of parameter space classification in the case of mixture of Gaussian distributions.
 
%

%
%

As we have discussed before, we have introduced 7 families of pulsars namely rotationally powered pulsars (RPPs) that involve millisecond pulsars, rotating radio transients (RRATs), gamma-ray pulsars (GRPs) as well as conventional radio pulsars, 
magnetars (AXPs, SGRs), X-ray dim INS (XDINSs) and central compact objects (CCOs).
As seen in \autoref{fig:dpgmm_pp1}, DPGMM identifies six clusters for the value of the concentration parameter $\dparam=10^{-5}$. 
In this section, we analyse whether these six clusters are linked to any parameters of pulsars other than $\per$ and $\perdot$ (e.g.\ space velocity or surface temperature) that could attribute further meaning to the different clusters.
%

\begin{table}
\centering
\begin{tabular}{@{}cccccccc@{}}
\toprule
 & \multicolumn{6}{c}{Cluster} &  \\ \midrule
Pulsar Family & \first & \second & \third & \fourth & \fifth & \sixth & \textbf{Total} \\
\midrule
Normal & 137 & 73 & 703  & 906 & 142 & 21 & \textbf{1982} 
\\
GRP & 38 & 2 & 0 & 15 & 62 & 0 & \textbf{117} 
\\
RRAT & 0 & 0 & 21  & 12 & 1 & 1 & \textbf{35} 
\\
AXP & 0 & 0 & 4 & 0 & 0 & 10 & \textbf{14} 
\\
SGR & 0 & 0 & 0 & 0 & 0 & 8 & \textbf{8} 
\\
XDINS & 0 & 0 & 6 & 0 & 0 & 1 & \textbf{7} 
\\
CCO & 0 & 0 & 3 & 0 & 0 & 0 & \textbf{3} 
\\ \midrule
\textbf{Total} & \textbf{175} & \textbf{75} & \textbf{737} & \textbf{933} & \textbf{205} & \textbf{41} & \textbf{2166} \\ 
\bottomrule
\end{tabular}
\caption{The distribution of the pulsar families over clusters. 
}
\label{table:family_distv2}
\end{table}

\subsection{Pulsar distribution over the clusters}
\label{subsec:pulsar_dist_over_clusters}

The distribution of pulsar types over DPGMM clusters is given in \autoref{table:family_distv2}. 
Accordingly, the most of the `non-recycled' RPPs detected in the radio band are located in the \third, \fourth\ and \fifth\ clusters. 
The \first\ and the \second\ regions contain mostly millisecond pulsars. 
It is significant that millisecond pulsars are separated into two subclasses by DPGMM  as in the case of GMM \citep{lee+12}. 
In \S~\ref{sec:discuss}, we discuss the possible origin of this division.
The \second\ region is very large and extends beyond the pulsar death line. 
The \third\ region contains most of the RRATs and XDINSs, all CCOs, low-B magnetars (J1647--4552, J0418+5732, J2301+5852 and J1822--1604) and those RPPs close to the death line. 
There is again a \fourth\ region populated by RPPs with a narrow $\perdot$ range. 
About $10\%$ of GRPs overflow to this region. 
The \fifth\ region mostly contains `non-recycled' GRPs while the recycled ones are in the \first\ cluster. 
The \sixth\ region is predominantly occupied by 'High-B` AXPs and SGRs excluding the low-B magnetars. 
Moreover, most of the pulsars are grouped in the \third\ and \fourth\ clusters that are located at the center of the diagram. 
There are fewer pulsars in other clusters relatively.
 
\subsection{Effect of binary companion for recycled pulsars}
\label{subsec:binary_companion_dist_over_clusters}

The distribution of binary pulsars over clusters is given in \autoref{table:binary_pulsar_dist}. 
Accordingly, the binary pulsars are heavily grouped in the \first\ and the \second\ clusters where millisecond pulsars `live'. 
This, of course, is understood through the ``recycling scenario'' \citep{alp+82} that address the very existence of these objects by spin-up of the neutron star through long-term accretion of matter and angular momentum from a binary companion. 
It is interesting, however, that both GMM and DPGMM found the millisecond pulsars are distributed into two clusters. 
What could be the underlying astrophysical distinction between these two clusters?

\begin{figure}
  \includegraphics[width=\linewidth]{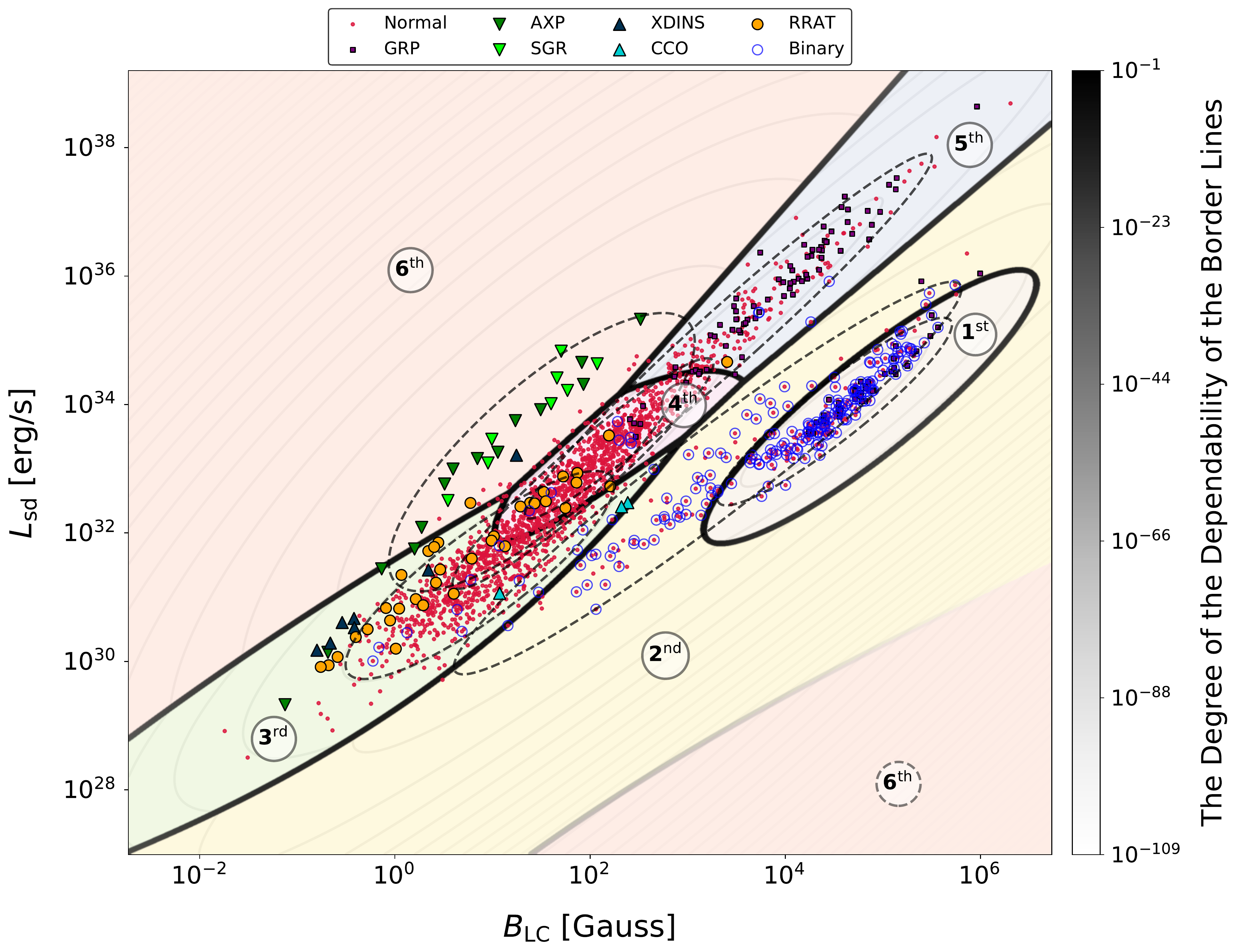}
\vskip -2mm
    \caption{
    The $\bfieldLC-\sdpow$ parameter space classified by DPGMM.
    }
\label{fig:BLC-Edot}
\end{figure}

Almost all millisecond GRPs are located in the \first\ cluster as seen in \autoref{fig:dpgmm_pp1} and in \autoref{table:binary_pulsar_dist}. 
This is obviously due to the higher rotational power, $\sdpow = 4\pir^2 I \perdot/\per^3$, of this cluster as well as the magnetic field strength at the light cylinder radius $\pradLC = c\per/2\pir$ given by 
\begin{equation}
\bfieldLC = \sqrt{\frac{24\pir^4 I \perdot}{c^3 \per^5}} \enskip .
\end{equation}
This can more readily be inferred from \autoref{fig:BLC-Edot} where DPGMM classification of $\bfieldLC-\sdpow$ parameter space is illustrated.  

As $\lbrace \sdpow, \bfieldLC \rbrace$ are the linear functions of $\lbrace \per, \perdot \rbrace$ in logarithmic scale such that
	\begin{align}
	\log(\bfieldLC) &\equiv \frac{1}{2}\log(\perdot) - \frac{5}{2}\log(\per) 
	\enskip ,  \\
	\log(\sdpow) &\equiv \log(\perdot) - 3\log(\per) 
	\enskip , 
	\end{align}	
then the data matrix $\mybold{Y} = [\log\mybold{B}_{\rm LC} \,\, \log\mybold{L}_{\rm sd}]$ is also distributed with the mixture of Gaussians as is the original data $\dset = [\log\mybold{\per} \,\, \log\mybold{\perdot}]$. 
This is true for any data matrix $\mybold{Y}$ which is linearly related with $\dset$.  
Therefore, an analysis in the $\bfieldLC-\sdpow$ parameter space is equivalent to the analysis in the $\pspace$ diagram. 
For this reason, we applied the same DPGMM ($\maxclust=10$, $\dparam=10^{-5}$) to the $\bfieldLC-\sdpow$ parameter space. 
The obtained pulsar classification in this plane is almost same as the classification in $\pspace$ ~parameter space as should be. 
As seen in \autoref{fig:BLC-Edot}, $\gamma$-ray emission can not be the underlying astrophysical cause for the existence of two separate clusters for millisecond pulsars; it will rather be a consequence.

\begin{table}
\centering
\begin{tabular}{@{}cccc@{}}
\toprule
\textbf{Cluster} & \textbf{Normal} & \textbf{GRPs} & \textbf{Total} \\ \midrule
\first & 103 & 32 & \textbf{135} \\
\second & 51 & 0 & \textbf{51} \\
\third & 13 & 0 & \textbf{13} \\
\fourth & 6 & 0 & \textbf{6} \\
\fifth & 3 & 1 & \textbf{4} \\
\sixth & 0 & 0 & \textbf{0} \\ \midrule
\textbf{Total} & \textbf{176} & \textbf{33} & \textbf{209} \\ \bottomrule
\end{tabular}
\caption{
The distribution binary pulsar families over clusters. 
}
\label{table:binary_pulsar_dist}
\end{table} 

\begin{table}
\centering
\begin{tabular}{@{}cccccc@{}}
\toprule
\textbf{Cluster} & \textbf{MS} & \textbf{NS} & \textbf{CO} & \textbf{HE} & \textbf{UL} \\ \midrule
\first & 9 & 0 & 15 & 77 & 23 \\
\second & 0 & 14 & 17 & 12 & 4 \\
\third & 0 & 2 & 3 & 4 & 2 \\
\fourth & 4 & 1 & 1 & 0 & 0 \\
\fifth & 2 & 1 & 0 & 1 & 0 \\
\sixth & 0 & 0 & 0 & 0 & 0 \\ \midrule
\textbf{Total} & \textbf{15} & \textbf{18} & \textbf{36} & \textbf{94} & \textbf{29} \\ \midrule
\end{tabular}
\caption{The distribution of the companion types of binary pulsars over clusters. Companion types --- main-sequence (MS), neutron star (NS), carbon-oxygen (or ONeMg) white dwarf (CO), helium white dwarf (HE), and ultra-light companion or planet (UL) --- are obtained from Australia Telescope National Facility (ATNF) Pulsar Catalog (http://www.atnf.csiro.au/research/pulsar/psrcat/).}
\label{table:binary_companion_dist}
\end{table}

The companion types of some binary pulsars have been detected and the distribution of these companions over the clusters is shown in \autoref{table:binary_companion_dist}. 
Accordingly, there is a remarkable distinction between the properties of the companion objects in the two groups indicating to different astrophysical origins and recycling history. 
The \first\ cluster is populated by millisecond pulsars with main sequence (MS), helium white dwarf (HE), carbon-oxygen white dwarf (CO) and ultra-light (UL) companions while lacking any neutron star companions. 
The \second\ cluster is populated by those with neutron star (NS) and white dwarf companions (CO and He), while 
having only 4 UL and no MS companions. 
\citet{lee+12} attributes the identification of two separate clusters to the chemical composition of the companions.  
This implies that millisecond pulsars in the \second\ cluster systematically had larger mass companions that lived shorter. 
Their larger periods (hence the smaller rotational power and lack of gamma-emission) can be attributed to the shorter recycling history they suffered. 
It is remarkable that DPGMM distinguishes these two groups only through their spin parameter distributions.

\subsection{Characteristic age and magnetic field}
\label{subsec:cage_bfield}

We calculated the characteristic age, $\cage \equiv \per/2\perdot$, and magnetic dipole field strength as inferred under the assumption of a orthogonal rotating magnetic dipole model
\begin{equation}
\bfield_{\rm d} = \left( \frac{ 3 c^3 I}{8\pir^2 R^6} \right)^{1/2} \sqrt{\per\perdot} 
\label{eq:Bd}
\end{equation}
of each cluster where we assumed $I=10^{45}\,{\rm g\,cm^2}$ and $R=10^6\,{\rm cm}$ for the moment of inertia and the radius of neutron stars, respectively. 
These parameters calculated based on the mean value ($\gmean[i] = \lbrace \per_i, \perdot_i \rbrace$) of corresponding Gaussian distribution. Therefore the results given in \autoref{table:derived_comp_params} can be assumed as the average values for each cluster. 
The table is arranged such that the age of cluster is increased with the cluster order as well as the magnetic dipole strength.
Hence, the \first\ cluster is the youngest one with the highest dipole field where the \sixth\ cluster is the oldest one with the lowest dipole field. 
Beside that, the millisecond pulsars in the \second\ cluster have an order of magnitude larger magnetic dipole field strengths compared to those in the \first\ cluster. 
According to our discussion in the previous subsection, stronger fields can be understood as a consequence of the shorter `recycling' episode these systems passed through.

\begin{table}
\centering
\begin{tabular}{@{}ccc@{}}
\toprule
\textbf{}    & \textbf{Characteristic Age} & \textbf{Surface Magnetic Flux} \\ 
\textbf{Cluster}  & \textbf{(yr)} & \textbf{(Gauss)} \\ \midrule

\first  & $4.49 \times 10^9$                      & $2.52 \times 10^8$                                   \\
\second  & $1.35 \times 10^9$                      & $2.96 \times 10^9$                                   \\
\third  & $3.06 \times 10^7$                      & $7.04 \times 10^{11}$                                \\
\fourth  & $2.70 \times 10^6$                      & $1.28 \times 10^{12}$                                \\
\fifth  & $1.15 \times 10^5$                      & $2.21 \times 10^{12}$                                \\
\sixth  & $6.75 \times 10^4$                      & $3.35 \times 10^{13}$                                \\ \midrule
\end{tabular}
\caption{The characteristic age (year) and dipole magnetic field strength (Gauss) of clusters.}
\label{table:derived_comp_params}
\end{table}

\subsection{Transverse velocity} 
\label{subsec:tvel}

\begin{figure}
  \includegraphics[width=\linewidth]{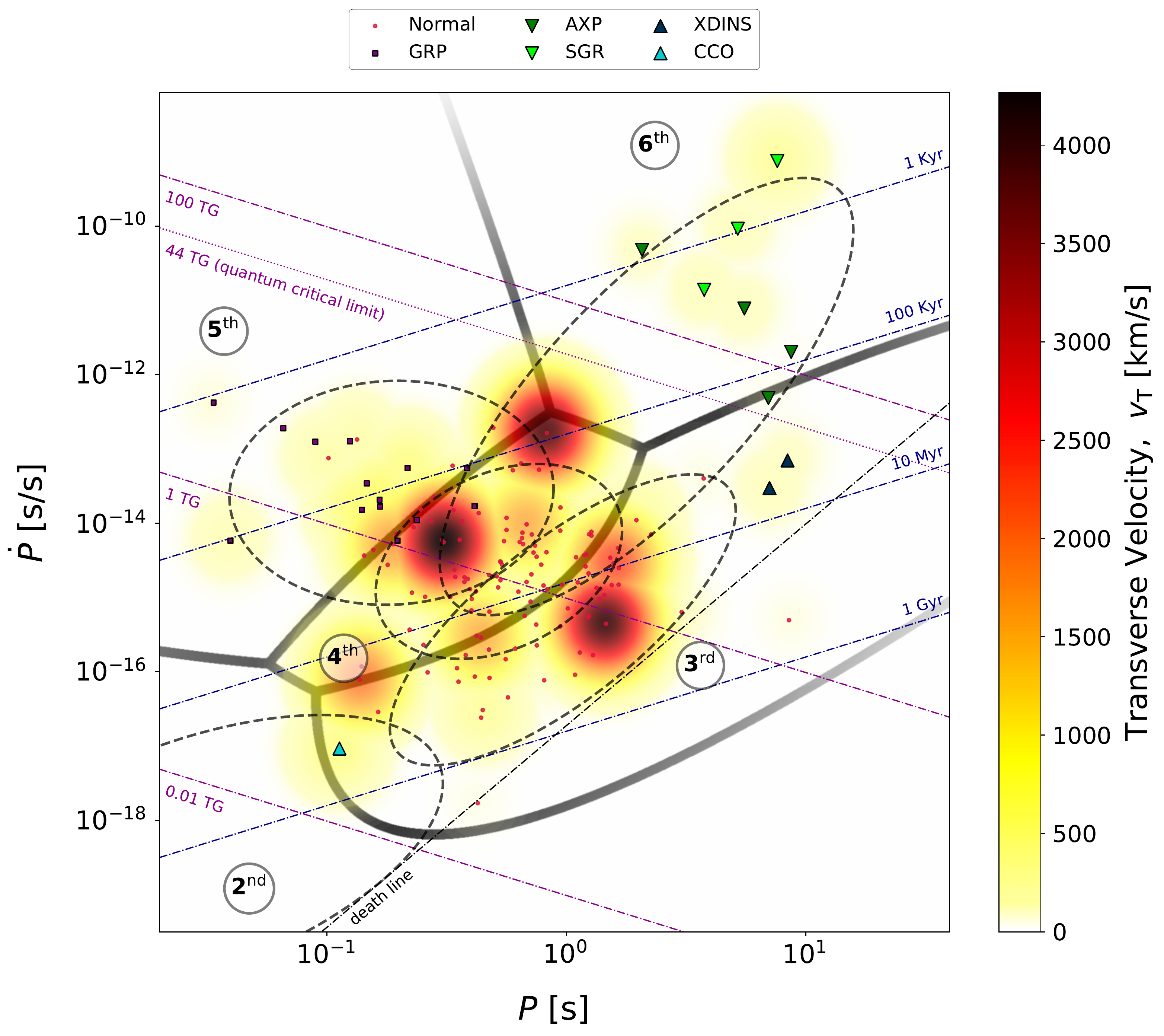}
\vskip -2mm
  \caption{ 
The transverse velocity of non-recycled pulsars (the \third, \fourth, \fifth\ and \sixth\ clusters).}
  \label{fig:fig_tvel}
\end{figure}

\begin{table}
\centering
\begin{tabular}{@{}ccrrrr@{}}
\toprule
            & \multicolumn{5}{c}{\textbf{Transverse Velocity $\tvel$ (km~s$^{-1}$)}}                                                                                                        \\ \midrule
\textbf{Cluster} & \textbf{$N_{\tvel}$} & \multicolumn{1}{c}{\textbf{$\tvelM$}} & \multicolumn{1}{c}{\textbf{$\tvelWM$}} &\multicolumn{1}{c}{\textbf{$\tvelS$}} &
\multicolumn{1}{c}{\textbf{$\tvelWS$}}  
\\ \midrule
\third & 50 & 335.5 & 363.3 & 625.4 & 677.4  \\
\fourth & 71 & 470.0 & 373.6 & 753.6 & 570.1 \\
\fifth & 20 & 302.6 & 359.1 & 248.8 & 265.4  \\
\sixth & 6  & 252.2 & 231.5 & 141.0 &  79.7 \\
\bottomrule
\end{tabular}
\caption{Statistics of transverse velocity $\tvel$ (km~s$^{-1}$) of non-recycled pulsars (the \third, \fourth, \fifth\ and \sixth\ clusters). Here $N_{\tvel}$ stands for the number of pulsars with the measured speed in each cluster. $\tvelM$, $\tvelWM$, $\tvelS$ and $\tvelWS$ stand for mean, weighted mean, standard deviation and weighted standard deviation of the transverse speed, respectively.}
\label{table:tbl_tvel}
\end{table}

According to the field burial model \citep{mus95,you95,gep+99,ho11}, an initial, brief but intense, fallback accretion episode \citep{che89} following the supernova explosion is important in shaping the magnetic fields of nascent pulsars. 
The amount of matter accreted determines how much the initial magnetic field is buried under the crust and thus the time-scale of diffusion of the field to the surface. 
Space velocities of pulsars due to the kick they receive during their birth may thus determine how much mass the neutron star can accrete, as the accretion rate is inversely proportional to the cube of the velocity in Bondi-Hoyle accretion. 
The field diffusion time-scale as inferred from the measured braking indices of young pulsars and the space velocities are inversely correlated \citep{gun13,eks17} as expected from the field-burial scenario. 
The question then naturally arises whether the space velocity of pulsars have any role in the observed diversity of pulsars.

In order to check this possibility we showed the transverse velocity of non-recycled pulsars (i.e.\ those in the \third, \fourth, \fifth\ and \sixth\ clusters in the $\pspace$ diagram) as seen in \autoref{fig:fig_tvel}. 
The figure is colored depending on the magnitude of the velocity so that the color changes with the magnitude from yellow to red sequentially. 
The transverse velocity statistics of same clusters are given in \autoref{table:tbl_tvel}. 
Since the probability that each pulsar is a member of the cluster to which they belong is different, we calculated the weighted statistics of their transverse velocities by using their joint probability values (see equation \eqref{eq:joint_probability}) with respect to corresponding Gaussian distributions as weights.  
Accordingly, the weighted standard error is very large due to the low number of space velocity measurements in each group and the results may not be statistically significant. 
We see, however, that the space velocities of pulsars in the \fourth\ cluster is the largest. 
We also see that the average speeds of pulsars in the \sixth\ cluster is about half of those in the \fourth\ cluster. 
The velocity data was collected from the sources in which the spin parameters of pulsars were taken as specified in \S~\ref{subsec:data_collection}.

\subsection{Blackbody temperature}
\label{subsec:ktbb}

\begin{figure} 
  \includegraphics[width=\linewidth]{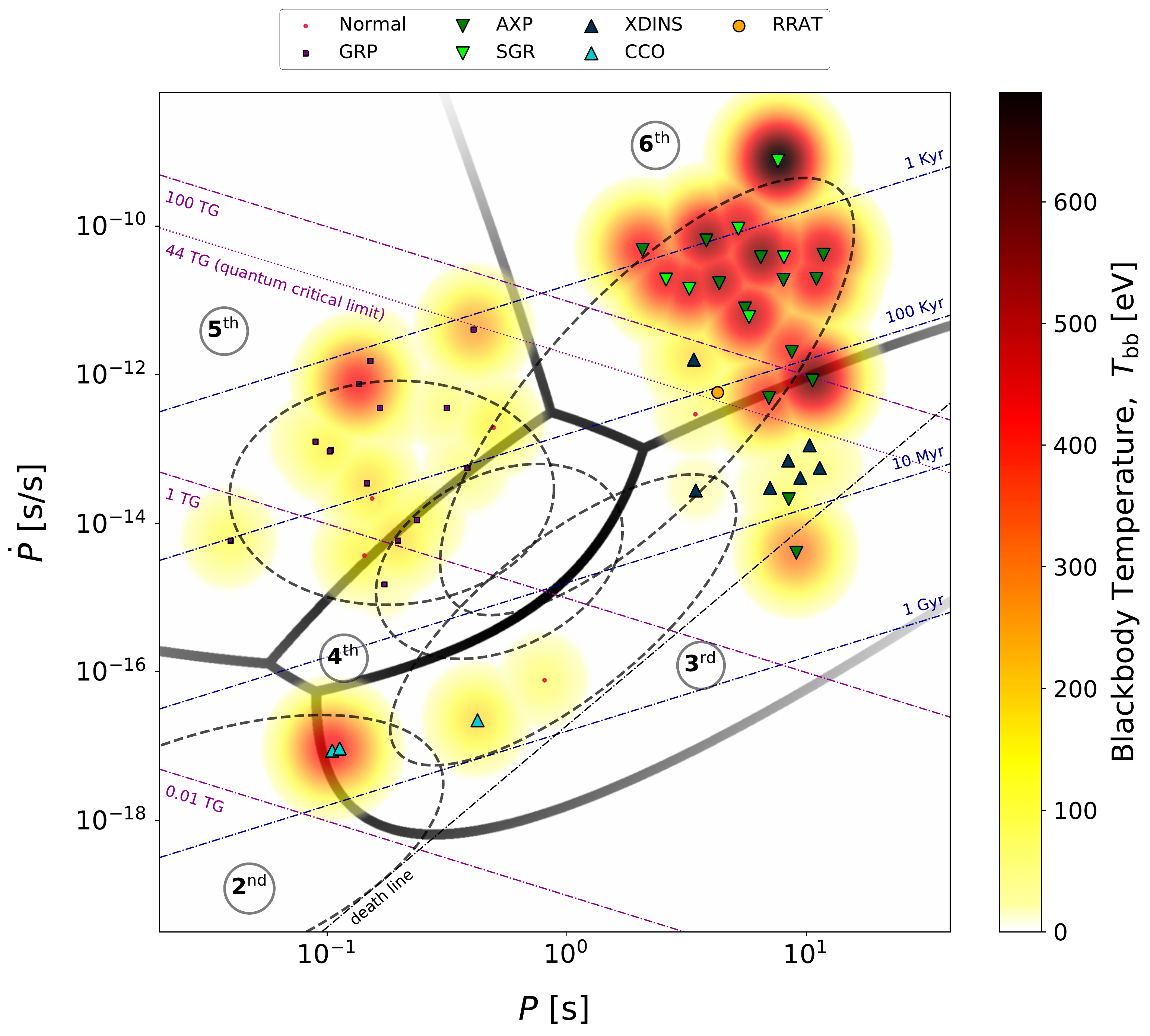}
  \vskip -2mm
  \caption{
The blackbody temperature (eV) of non-recycled pulsars (the \third, \fourth, \fifth\ and \sixth\ clusters).}
  \label{fig:fig_ktbb}
\end{figure}

In \autoref{fig:fig_ktbb}, the blackbody temperature of non-recycled pulsars are illustrated. The statistics of black body temperature of these clusters are given in \autoref{table:tbl_ktbb}. 
The weighted statistics are computed in the same way described in the previous section.
According to these results, the temperature of the \sixth\ cluster where magnetars are frequently hosted in is almost twice as large as temperatures of other clusters.  
In other words, AXP/SGRs are the hottest pulsar type. 
However, the results may not be statistically significant due to the small sample size. 
The blackbody temperatures of these pulsars are acquired from the sources mentioned in \S~\ref{subsec:data_collection} and also from \citet{gon+07,kasp04,webb+04,hu+17,li+05,chan+11,cara+10,mare+11,guver+12,poss+12}.

\begin{table}
\centering
\begin{tabular}{@{}ccrrrr@{}}
\toprule
 & \multicolumn{5}{c}{\textbf{Blackbody temperature $\ktbb$ (eV)}}                                                                                                       \\ \midrule
\textbf{Cluster} & \textbf{$N_{\ktbb}$} & \multicolumn{1}{c}{\textbf{$\ktbbM$}} & \multicolumn{1}{c}{\textbf{$\ktbbWM$}} & \multicolumn{1}{c}{\textbf{$\ktbbS$}} &
\multicolumn{1}{c}{\textbf{$\ktbbWS$}}  
\\ \midrule
\third & 14 & 219.2 & 138.2 & 175.0 & 83.7 \\
\fourth &  3 & 128.7 & 126.3 & 67.7 & 55.4   \\
\fifth & 14 & 172.0 & 157.5 & 84.4 & 51.0   \\
\sixth & 19 & 395.5 & 334.1 & 166.5 & 171.3 
\\ \bottomrule
\end{tabular}
\caption{Statistics of blackbody temperature (eV) of non-recycled pulsars (the \third, \fourth, \fifth\ and \sixth\ clusters). Here $N_{\ktbb}$ stands for number of pulsars with the measured temperatures. $\ktbbM$, $\ktbbWM$, $\ktbbS$ and $\ktbbWS$ stand for mean, weighted mean, standard deviation and weighted standard deviation of temperature, respectively.}
\label{table:tbl_ktbb}
\end{table}

\section{Discussion}
\label{sec:discuss}

We have examined the pulsar population with an unsupervised machine learning algorithm, Dirichlet process Gaussian mixture model (DPGMM). 
We confirmed the earlier result obtained with Gaussian mixture model (GMM) by \citet{lee+12} that the millisecond pulsar population has two clusters and that the normal pulsar population consists of four clusters. 

We have considered possible hidden parameters (space velocity and surface temperature) as possible underlying cause of the distinct clusters. 
We have first tried DPGMM on higher dimensional planes, e.g.\  3-D planes ($\per-\perdot-\tvel$ and $\per-\perdot-\ktbb$) and 4-D plane ($\per-\perdot-\tvel-\ktbb$). 
We have found that the results for the higher dimensions were unreliable due to the lack of the data instances available for the time being.
We have then examined the transverse velocity and the blackbody temperature statistics for each cluster.
We have found that the \fourth\ cluster has about twice the average space velocity compared to the slowest cluster, the \sixth\ one, (see \autoref{fig:fig_tvel} and \autoref{table:tbl_tvel}) providing a marginal evidence that space velocities of young neutron stars could play a role in their astrophysical manifestations by controlling the total mass accreted during post-supernova fallback and the depth of field burial \citep{gun13,eks17}. 
This indicates the pulsars in the \fourth\ cluster suffered less supernova fallback accretion and less field burial. 
Indeed the average magnetic field of pulsars in the \fourth\ cluster are similar to those in the \fifth\ cluster and are twice of those in the \third\ cluster. 
If the magnetic fields of pulsars decay as they age within a timescale less than a million years, the difference in the magnetic fields of the \fourth\ and \third\ clusters could be also attributed to the former systems being younger and subject to less field decay. 
Yet it is unlikely that the magnetic fields of typical pulsars decay in such short timescales favouring the idea that the space velocities could be the underlying reason for the difference in the magnetic fields of the \fourth\ and \third\ clusters.

We note that the very high space velocities of CCOs imply that they are not likely to suffer very strong post-supernova fallback accretion. 
This does not neatly fit into the `field burial scenario' \citep{mus95,gep+99} as the small dipole fields of these objects, in this model, are attributed to very intense fallback accretion. 
This may indicate the role of the initial magnetic field and spin period as other parameters controlling the amount of the accreted fallback mass: The low magnetic fields and relatively slow initial periods of CCOs could allow them to accrete large amount of mass to bury their fields in spite of their large space velocities. 
Another solution high-speed-low-field puzzle could be that their dipole fields are in fact very strong but their rotation and spin axis are aligned to a very high precision such that the cluster of their dipole moment perpendicular to the rotation axis inferred from spin-down is so small. 
This idea, however, would work only in the vacuum dipole model i.e.\ if a corotating plasma does not exist around CCOs; Magnetohydrodynamic (MHD) dipole model \citep{spi06} require non-vanishing spin-down even in the aligned case. 
This idea is worth considering given that only three of the CCOs show pulsations while there is evidence for strong magnetic fields \citep{sha+12}.

It is remarkable that the speed of objects in the \sixth\ group is `typical'. 
This group involves AXP/SGRs which according to the magnetar model should have large space velocities, $\sim 1000$~km~s$^{-1}$ due to the ``rocket effect'' \citep{dun92,tho93}.
This may be because it is their super-strong quadrupole fields, rather than the dipole fields, that distinguishes these objects from the rest of the isolated pulsars. 
The higher spin-down rates of these objects could then be attributed to additional torques, contributing to the magnetic dipole torque, due to winds \citep{har+99,tho+00} or supernova fallback discs \citep{wan+06,ert+07}. 

The surface temperatures of pulsars in the \sixth\ cluster are found to be relatively large (see \autoref{fig:fig_ktbb} and \autoref{table:tbl_ktbb}). 
This is an expected result as this cluster involves AXP/SGRs which, according to magnetar model, have decaying fields. 
This also is expected according to the fallback disc model as a result of accretion \citep{alp01,cha+00}. 
A critical discrimination between the two models would be to compare the surface temperatures of HBPSR's which presumably are not accreting objects, with those of AXP/SGRs. 
The upper limits on the temperatures of high magnetic field pulsars PSR B0154+61 \citep{gon+04}, J1814-1744 and J1847-0130 \citep{kea+13} imply that they are not significantly hotter than pulsars with lower magnetic fields thus raising the question why the fields of these objects do not decay and contribute to the X-ray luminosity. 
Yet the temperature of B1509-58 \citep{hu+17} lies between the temperatures of AXP/SGRs and typical RPPs, and the temperature of PSR J1119--6127 \citep{gon+05} is among the highest measured from typical RPPs. 
The observed pulsed fraction of this source is also higher than those of typical RPPs \citep{gon+05} possibly indicating the role of strong fields and unlikely to be related to the presence of a putative fallback disc. 
Also, the recent analysis \citep{cer+19} shows that the noise strength and X-ray luminosity  magnetars lacks the correlation known to exist in accreting X-ray pulsars thus suggesting that they are not accreting.


If cluster regions do indeed have anything to do with the evolutionary connections among pulsar families, the \fifth\ and the \sixth\ regions may be considered as two different origins for young neutron stars on the $\pspace$ diagram. 
A nascent neutron star may be born as a Crab-like GRP ($\per_0\sim 10$ ms, $\perdot_0\sim 10^{-12}$~s~s$^{-1}$) or as a magnetar ($\per_0\sim 0.5$~s, $\perdot_0\sim 10^{-8}$~s~s$^{-1}$) assuming the field is not produced by the initial dynamo effect, but flux conservation \citep{fer06} depending on its spin parameters, respectively. 
The classification presented here clearly relates low-B magnetars with XDINSs and RRATs and implies they could descend from magnetars by field decay. 
Yet statistics suggests magnetars are much less common objects compared to XDINSs and RRATs. 
This evolutionary scenario could, however, be validated only if magnetar stage is very short or there are many undiscovered transient magnetars. 
The recent identification of the high energy component in XDINS \citep{yon+19} indeed suggests that magnetars and XDINSs are evolutionary linked.

There are, of course, strong selection effects on the representation of pulsars on the $\pspace$ diagram. 
CCOs, RRATs and XDINSs should be much more common in the galaxy than they are detected. 
Yet it is not possible to do the similar analysis with a `flux limited sample' at present as such filtering of data would result with much less instances of data than the model we employed, DPGMM, requires. 
Given that many types of pulsars are lower-represented in our sample, it may be argued that some of the clusters discovered by DPGMM are artefacts of describing an intrinsically non-Gaussian distribution. 
Yet, DPGMM classification we present here broadly coincides with the existing astrophysical classification despite it is an unsupervised model.

In summary, we have shown that DPGMM, an unsupervised machine learning model, identifies 6 pulsar classes: two for recycled millisecond pulsars corresponding to different accretion histories; one roughly coinciding XDINSs, low-B magnetars and RRATs, but involves many RPPs as well; one involving the bulk of RPPs and some RRATs; one involving the very young Crab-like RPPs including the bulk of the GRPs; and finally one that involves the AXP/SGRs and HBPSRs. 

\section*{Acknowledgements}

KYE acknowledges support from TUBITAK, the national foundation of science in Turkey, with the project number 118F028.
We thank M.\ Ali Alpar and Erbil G{\"u}gercino{\u g}lu for useful discussion.




\bibliographystyle{mnras}
\bibliography{refs,grp,ml} 


%
%
%
%
%

\bsp	
\label{lastpage}
\end{document}